\documentclass[journal]{IEEEtran}

  \IEEEpubid{
  \begin{minipage}{\textwidth}
  \centering
  \footnotesize
  \copyright~2026 IEEE. All rights reserved, including rights for text and data mining, and training of artificial intelligence\\
  and similar technologies. Personal use is permitted, but republication/redistribution requires IEEE permission.\\
  See https://www.ieee.org/publications/rights/index.html for more information.
  \end{minipage}
  }

\usepackage{stfloats}
\usepackage{verbatim}
\usepackage{cite}
\usepackage{amsmath,amssymb,amsfonts}
\usepackage{algorithmic}
\usepackage{graphicx}
\ifCLASSINFOpdf
  \DeclareGraphicsExtensions{.pdf,.png,.jpg}
\else
  \DeclareGraphicsExtensions{.eps}
\fi
\usepackage{textcomp}
\usepackage{xcolor}
\usepackage{amsmath}
\usepackage{dsfont}
\usepackage{booktabs}
\usepackage{silence}
\usepackage{subcaption}
\def\BibTeX{{\rm B\kern-.05em{\sc i\kern-.025em b}\kern-.08em
    T\kern-.1667em\lower.7ex\hbox{E}\kern-.125emX}}
\begin{document}

\title{\LARGE Deep Learning Based Channel Extrapolation for Dual-Band Massive MIMO Systems\\
}
\author{Qikai Xiao, Kehui Li, Binggui Zhou, Shaodan Ma,~\IEEEmembership{Senior Member,~IEEE}
\thanks{This work was supported in part by the Science and Technology Development Fund, Macau SAR 
under Grants 0114/2025/AMJ, 0020/2025/RIB1, 001/2024/SKL and 0002/2025/EQP, and in part by 
the Research Committee of University of Macau under Grant MYRG-GRG2025-00143-IOTSC. \textit{(Corresponding author: Binggui Zhou.)}}
\thanks{Qikai Xiao and Kehui Li contributed equally to this work.}
\thanks{
Qikai Xiao, Kehui Li, Shaodan Ma are with the State Key Laboratory of Internet of Things for Smart City and the Department of Electrical and Computer Engineering, University of Macau, Macao 999078, China (e-mails: mc45242@um.edu.mo; yc47997@um.edu.mo; shaodanma@um.edu.mo).}
\thanks{
Binggui Zhou is with the Department of Electrical and Electronic Engineering, Imperial College London, London, SW7 2AZ, U.K. (email: binggui.zhou@imperial.ac.uk).}
\thanks{
Digital Object Identifier 10.1109/LWC.2026.3689267
}
}
\maketitle
\IEEEpubidadjcol

\begin{abstract}
Future wireless communication systems will increasingly rely on the integration of millimeter wave (mmWave) and sub-6 GHz bands to meet heterogeneous demands on high-speed data transmission and extensive coverage. To fully exploit the benefits of mmWave bands in massive multiple-input multiple-output (MIMO) systems, highly accurate channel state information (CSI) is required. However, directly estimating the mmWave channel demands substantial pilot overhead due to the large CSI dimension and low signal-to-noise ratio (SNR) led by severe path loss and blockage attenuation. In this paper, we propose an efficient \textbf{M}ulti-\textbf{D}omain \textbf{F}usion \textbf{C}hannel \textbf{E}xtrapolator (MDFCE) to extrapolate sub-6 GHz band CSI to mmWave band CSI, so as to reduce the pilot overhead for mmWave CSI acquisition in dual-band massive MIMO systems. Unlike traditional channel extrapolation methods based on mathematical modeling, the proposed MDFCE combines the mixture-of-experts framework and the multi-head self-attention mechanism to fuse multi-domain features of sub-6 GHz CSI, aiming to characterize the mapping from sub-6 GHz CSI to mmWave CSI effectively and efficiently. The simulation results demonstrate that, across various antenna array scales and SNR levels, MDFCE achieves an excellent balance between performance and computational complexity compared with existing methods.
\end{abstract}

\section{Introduction}
\IEEEPARstart{F}{uture} wireless communication systems are expected to support heterogeneous requirements for both ultra-high-speed data transmission and wide-area coverage \cite{chen2024revolution}. Integrating diverse frequency bands, such as millimeter wave (mmWave) and sub-6 GHz bands, is a promising approach to achieving this goal \cite{gao2026aidriven}. To fully exploit mmWave bands, large-scale antenna arrays and massive subcarriers are usually deployed, and highly accurate mmWave channel state information (CSI) is required for precoder and decoder design \cite{alkhateeb2014mimo}, leading to substantial pilot training. Moreover, severe path loss and blockage attenuation lead to low signal-to-noise ratio (SNR), making direct mmWave channel estimation difficult. Compared to mmWave bands, sub-6 GHz bands operate at lower frequencies with smaller antenna dimensions and subcarriers, making sub-6 GHz CSI acquisition less pilot-intensive \cite{gao2021fusionnet} and less challenging because of lower path loss and blockage attenuation \cite{ma2023deep}. Although operating at different frequencies, mmWave and sub-6 GHz signals experience similar electromagnetic environments, leading to some common channel characteristics \cite{pasic2023statistical}. Based on these observations, recent works have explored extrapolating mmWave CSI from sub-6 GHz CSI to reduce the pilot overhead of direct mmWave CSI acquisition. The authors in \cite{pasic2024channel} proposed three phase-rotation-based methods that leverage sub-6 GHz CSI to estimate mmWave channels, thereby reducing in-band mmWave pilot overhead. However, these methods rely on specific channel-model assumptions and accurate knowledge of transmit-receive antenna-element distances, which limits their robustness and practical deployment. In addition, deep learning (DL) methods have demonstrated great potential for channel extrapolation because of the universal approximation capability of neural networks. The works \cite{zhou2025low} and \cite{gao2026enabling} explored Transformer-based multi-domain feature extraction from partial mmWave CSI and further leveraged generative artificial intelligence to perform in-band multi-domain channel extrapolation, thereby reducing channel estimation overhead. In addition, \cite{gao2026ssnet} developed a flexible self-supervised channel extrapolation framework for fluid antenna systems. Moreover, \cite{gao2021fusionnet,ma2023deep,li2024physicsenabled} investigated cross-band channel extrapolation. Specifically, \cite{gao2021fusionnet} and \cite{ma2023deep} focus on coarse-grained channel prediction, i.e., predicting optimal mmWave beam indices or spatial directions by fusing sub-6 GHz CSI with a few prior mmWave measurements, whereas \cite{li2024physicsenabled} targets the more challenging fine-grained task of full mmWave channel estimation, which aims to reconstruct the complete high-dimensional channel matrix.
\IEEEpubidadjcol
Nonetheless, the large frequency gap between the sub-6 GHz and mmWave bands makes cross-band CSI extrapolation highly nonlinear and intractable, making accurate channel extrapolation extremely challenging. Conventional methods rely heavily on strong model priors and therefore suffer from poor robustness. Although DL-based methods have demonstrated excellent performance through multi-domain feature extraction, they still face two critical limitations. First, advanced architectures such as Transformers and generative models incur high computational overhead, hindering deployment in resource-constrained systems. Second, existing cross-band studies remain predominantly confined to coarse-grained beam prediction. This coarse estimation cannot reconstruct the complete high-dimensional mmWave channel matrix and therefore lacks the precise amplitude and phase information required for advanced digital precoding. Therefore, direct fine-grained extrapolation from the sub-6 GHz to the mmWave band must balance robust multi-domain feature extraction with low computational complexity to unlock the full achievable rate of mmWave MIMO systems. To tackle these challenges, we propose the \textbf{M}ulti-\textbf{D}omain \textbf{F}usion \textbf{C}hannel \textbf{E}xtrapolator (MDFCE), a novel DL-based architecture for accurate and efficient mmWave channel estimation via cross-band channel extrapolation. The main contributions are summarized as follows:
\noindent\textit{1)} We employ multi-head self-attention (MHSA) and feed-forward networks (FFNs) in the MDFCE to extract spatial-frequency and spatial-delay domain features from sub-6 GHz CSI, enabling accurate low-overhead cross-band extrapolation to mmWave CSI. 
\noindent\textit{2)} We design a mixture-of-experts (MoE)-based multi-domain feature fusion architecture, where spatial-delay domain features adaptively fuse spatial-frequency domain features, enabling the network to capture diverse cross-band characteristics while significantly reducing complexity.
\noindent\textit{3)} Evaluated on DeepMIMO \cite{alkhateeb2019deepmimo}, the proposed MDFCE achieves a 4.44 dB gain over traditional least-squares (LS)-based mmWave channel estimation. Without providing mmWave band information, its MoE-based multi-domain fusion architecture further provides a 1.1 dB performance gain and a 1.33$\times$ faster inference speed over the Transformer-based network (TBN), while achieving a 9.64 dB gain over the state-of-the-art (SOTA) multimodal convolutional block attention module (MCBAM) with a simple design.
\section{System Model and Problem Formulation}
\begin{figure}[t]
    \centering
    \includegraphics[width=0.75\linewidth]{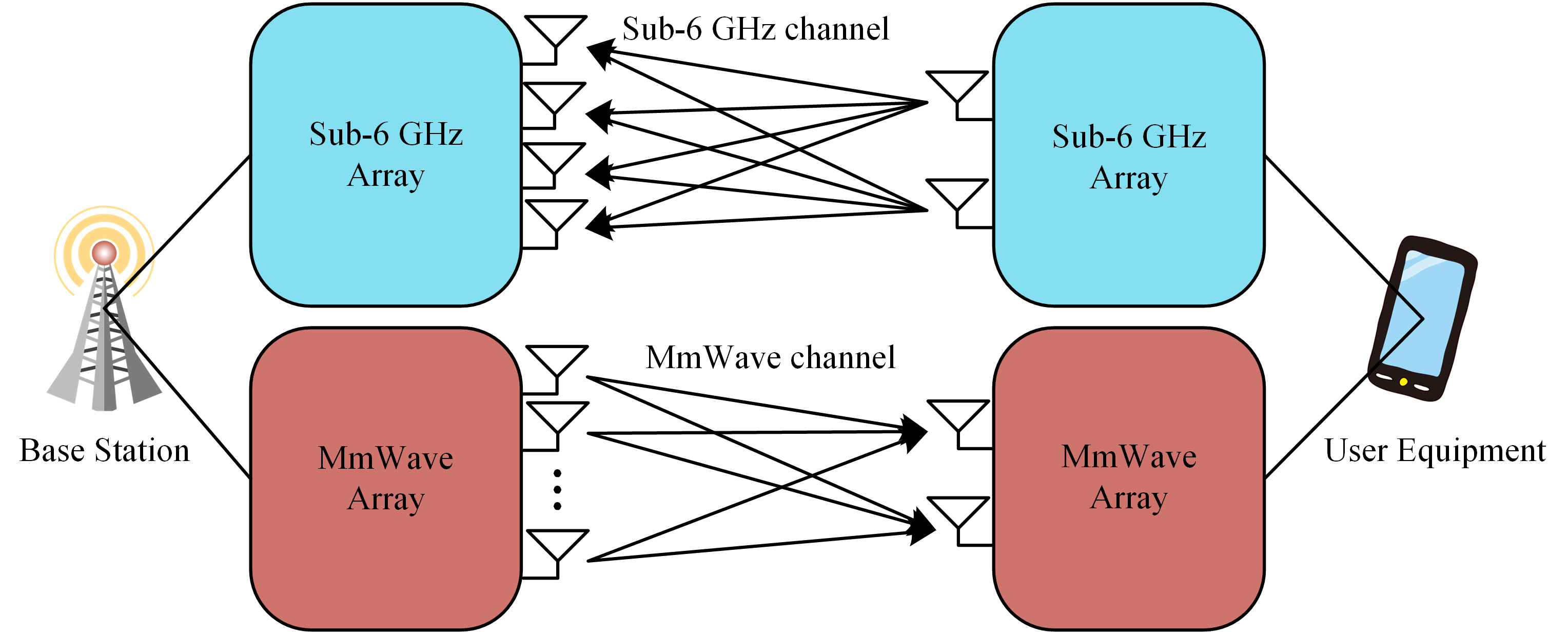}
    \caption{The dual-band massive MIMO system.}
    \label{fig:system-model}
\end{figure}
\subsection{System Model}
We consider a dual-band massive MIMO system, where sub-6 GHz and mmWave transceivers operate simultaneously. The uplink operates in the sub-6 GHz band with $M^s_{U}$ antennas at the transmitter and $M^s_{B}$ antennas at the receiver. The downlink operates in the mmWave band with $M^m_{B}$ antennas at the transmitter and $M^m_{U}$ antennas at the receiver. We assume that the transceivers at the two frequency bands are co-located at both the base station (BS) and the user equipment (UE), as shown in Fig. \ref{fig:system-model}. In addition, orthogonal frequency division multiplexing (OFDM) with $K^s$ subcarriers in the sub-6 GHz band and $K^m$ subcarriers in the mmWave band is adopted. The received signal at the $k$-th subcarrier in band $b\in\{s,m\}$ can be written as
\begin{equation}
\mathbf{Y}^{b}_{k}=\mathbf{H}^{b}_{k}\mathbf{X}^{b}_{k}+\mathbf{N}^{b}, \quad k=1,\dots,K^{b}.
\end{equation}
where $\mathbf{H}^{b}_{k} \in \mathbb{C}^{M^b_{B}\times M^b_{U}}$, $\mathbf{X}^{b}_{k} \in \mathbb{C}^{M^{b}_{U} \times M^{b}_{U}}$, and $\mathbf{N}^{b} \in \mathbb{C}^{M^b_{B}\times M^b_{U}}$ denote the CSI, transmitted signal matrix, and additive white Gaussian noise (AWGN), respectively, at the $k$-th subcarrier in band $b$.
By concatenating the per-subcarrier channels across all subcarriers, the spatial-frequency domain channel in band $b\in\{s,m\}$ can be written as
\begin{equation}
\mathbf{H}^{b}=[\mathbf{H}^{b}_1,\dots,\mathbf{H}^{b}_{K^b}],\quad
\mathbf{H}^{b} \in \mathbb{C}^{M^{b}_{B} \times (M^{b}_{U} \times K^{b})}.
\end{equation}
\subsection{Problem Formulation}
We assume that the whole sub-6 GHz channel $\hat{\mathbf{H}}^{s}$ has already been estimated at the BS using existing channel estimation methods \cite{zhou2024pay}. Our goal is to construct a mapping function $F_f$ that extrapolates the estimated sub-6 GHz channel $\hat{\mathbf{H}}^{s}$ to the mmWave channel $\hat{\mathbf{H}}^{m}$, thereby obtaining the mmWave channel while incurring only the pilot overhead of sub-6 GHz channel estimation. Since the sub-6 GHz channel may have much lower dimension than the mmWave channel and is much less susceptible to path loss and blockage attenuation, estimating the sub-6 GHz channel requires far lower pilot overhead than estimating the mmWave channel. The cross-band extrapolation problem can be formulated as 
\begin{align}
\min_{F_f} \quad &\mathcal{L}(\mathbf{H}^{m}, \hat{\mathbf{H}}^{m}), \notag \\
\text{s.t.} \quad [\mathrm{Re}(\hat{\mathbf{H}}^{m}), \mathrm{Im}(\hat{\mathbf{H}}^{m})]
&= F_f([\mathrm{Re}(\hat{\mathbf{H}}^{s}), \mathrm{Im}(\hat{\mathbf{H}}^{s})]).
\end{align}
where $\mathcal{L}(\cdot)$ denotes the loss function, e.g., a mean squared error (MSE) loss function. $\mathrm{Re}(\cdot)$ and $\mathrm{Im}(\cdot)$ represent the real and imaginary parts of the complex input, respectively.

However, the mapping function $F_f$ is highly nonlinear and intractable, which makes mathematically describing the mapping between sub-6 GHz and mmWave channels extremely difficult. To solve these issues, we design the lightweight MDFCE to learn this mapping efficiently by exploiting the nonlinear approximation capability of neural networks. 
\begin{figure*}[t]
    \centering
    \begin{subfigure}[t]{0.63\textwidth}
        \centering
        \includegraphics[width=\linewidth,keepaspectratio]{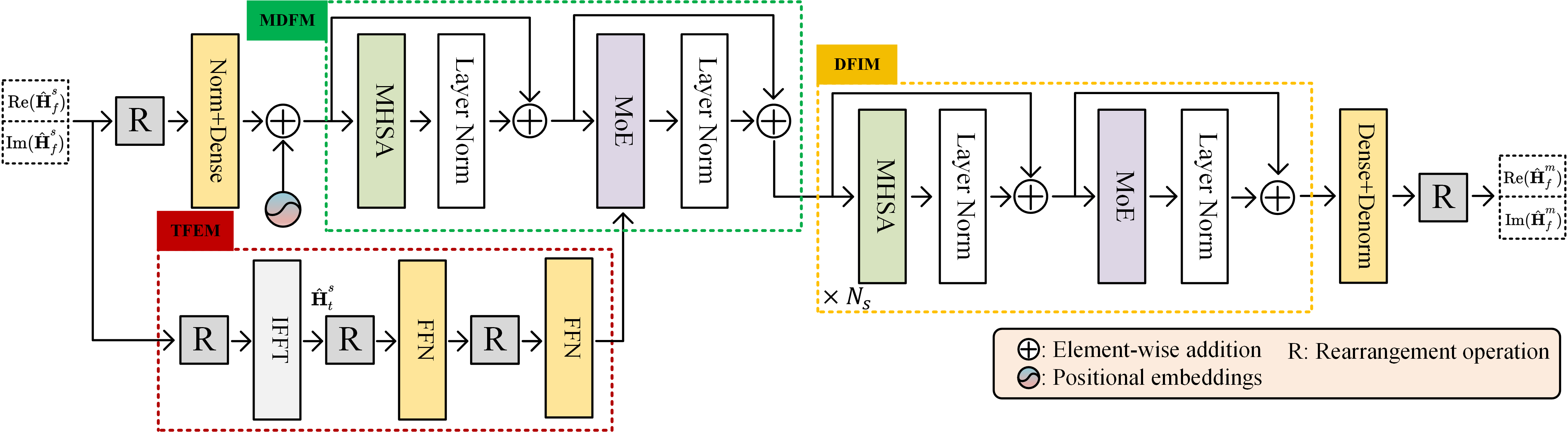}
        \caption{Overall MDFCE architecture}
        \label{fig:MDFCE}
    \end{subfigure}
    \hfill
    \begin{subfigure}[t]{0.17\textwidth}
        \centering
        \includegraphics[width=\linewidth,keepaspectratio]{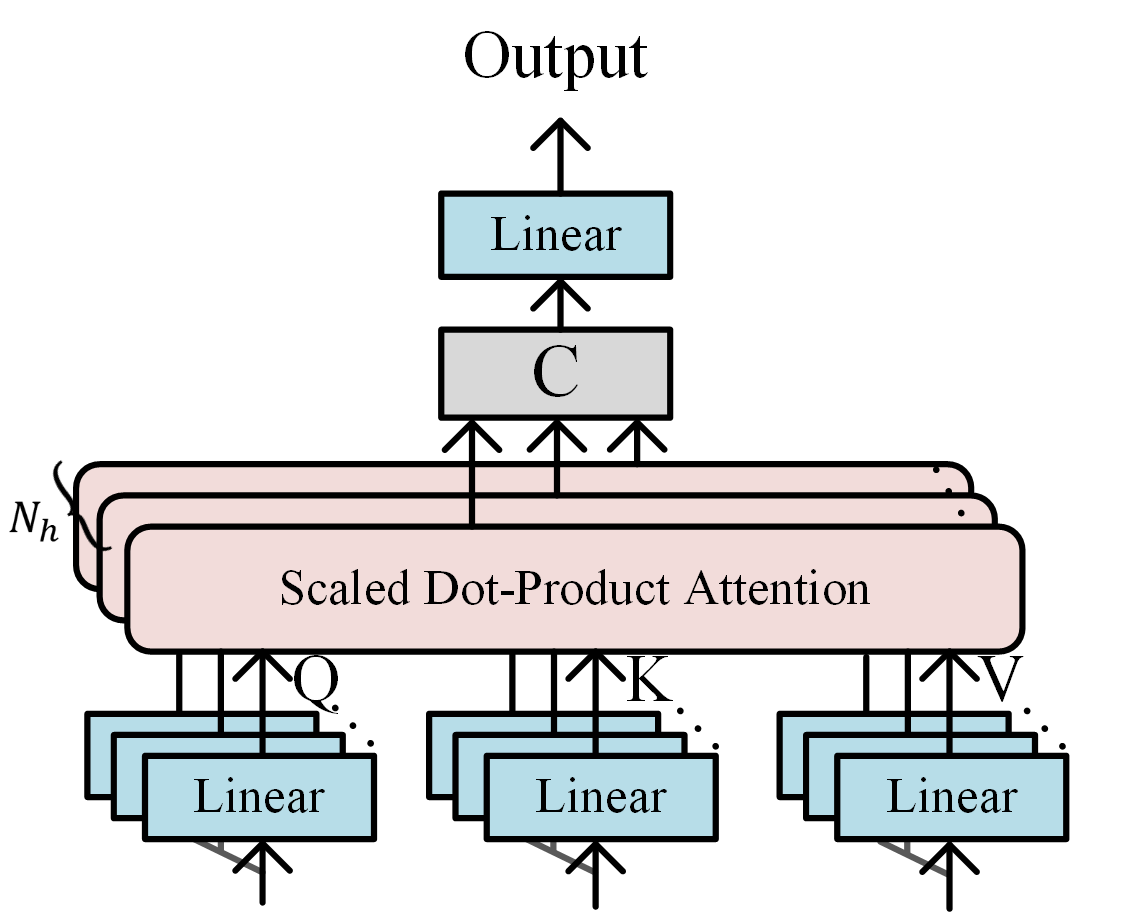}
        \caption{MHSA layer}
        \label{fig:MHSA}
    \end{subfigure}
    \hfill
    \begin{subfigure}[t]{0.17\textwidth}
        \centering
        \includegraphics[width=0.88\linewidth,keepaspectratio]{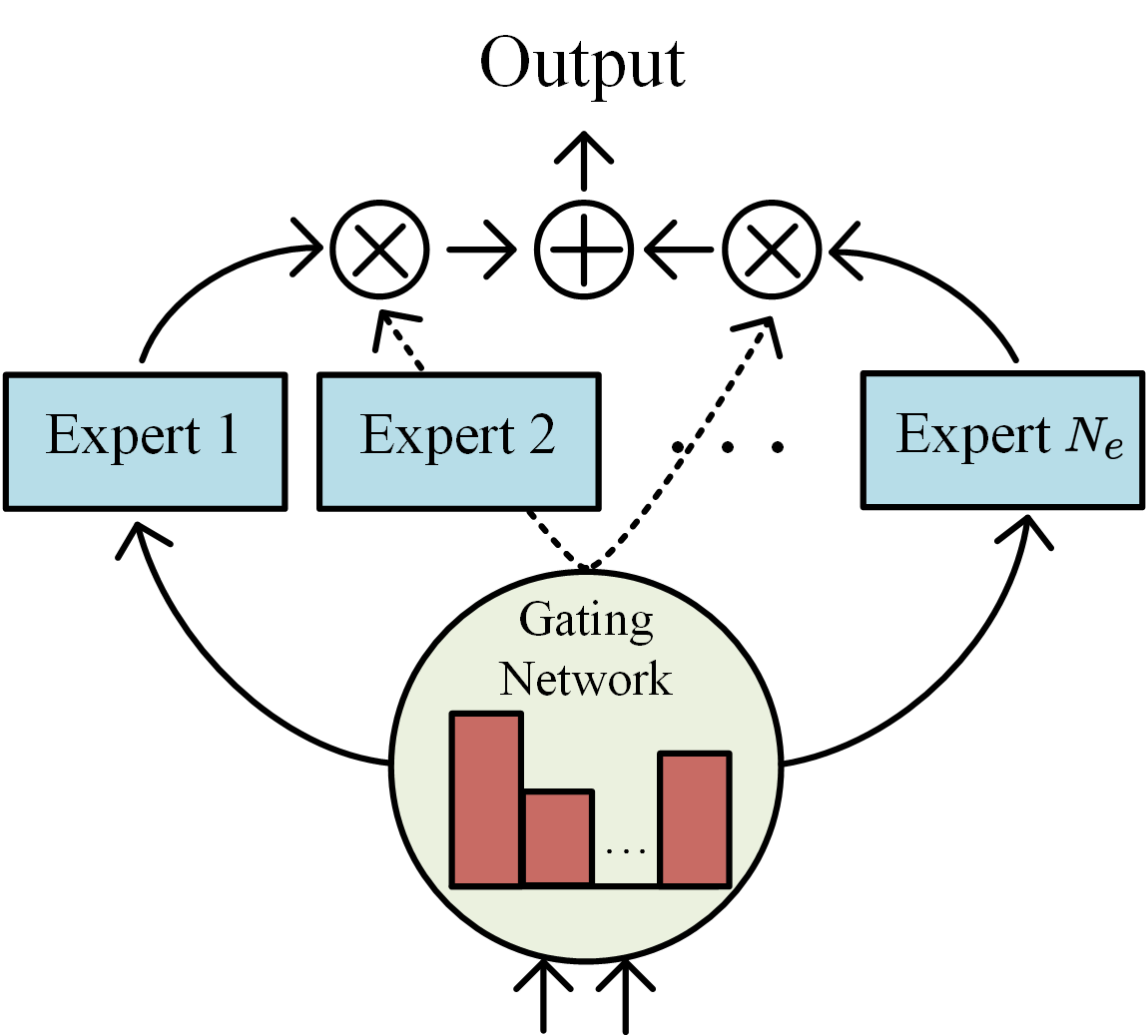}
        \caption{MoE layer}
        \label{fig:MoE}
    \end{subfigure}
    \caption{Architecture of the proposed MDFCE and illustration of key components in the MDFCE.}
    \label{fig:network_layers}
    \vspace{-8mm}
\end{figure*}
\section{Multi-Domain Fusion Channel Extrapolator}
To realize low-latency and accurate cross-band extrapolation, we propose a sparse MoE-based MDFCE (Fig. \ref{fig:MDFCE}), which jointly exploits spatial, frequency, and delay features of sub-6 GHz CSI to estimate mmWave CSI in an end-to-end manner.

\textbf{Overall Architecture}: The proposed MDFCE consists of three modules: the Time-delay Feature Extraction Module (TFEM), the Multi-Domain Fusion Module (MDFM), and the Deep Feature Interaction Module (DFIM). First, the lightweight TFEM transforms spatial-frequency sub-6 GHz CSI into the spatial-delay domain and generates a latent representation. Meanwhile, the MDFM extracts spatial-frequency features via MHSA \cite{vaswani2017attention}, and the latent representation generated by TFEM guides adaptive feature selection and fusion in the MoE layer \cite{shazeer2017outrageously}. Then, the DFIM refines the fused multi-domain representation through stacked MHSA and MoE layers to produce a latent embedding of the mmWave channel. Finally, an output layer maps this embedding back to the spatial-frequency domain to obtain the estimated mmWave CSI.

\textbf{TFEM}: To capture the delay domain characteristics of sub-6 GHz CSI, e.g., delay spread and multipath components, a lightweight TFEM is proposed. Specifically, the inverse fast Fourier transform (IFFT) is first applied to convert the spatial-frequency CSI \(\hat{\mathbf{H}}^s_f\) to spatial-delay CSI \(\hat{\mathbf{H}}^s_t\in \mathbb{C}^{M^{s}_{B}\times (M^{s}_{U}\times K^s)}\):
\begin{equation}
{\hat{\mathbf{H}}}_t^s = \text{IFFT}({\hat{\mathbf{H}}}_f^s),
\label{eq:IFFT}
\end{equation}
where \(\text{IFFT}(\cdot)\) denotes the IFFT operation. Subsequently, two feed-forward networks (FFNs) are employed to extract features and produce the latent representation \(\mathbf{X}_{t}^{s}{'}\in\mathbb{R}^{(M^{s}_{B} \times M^{s}_{U}) \times d_{re}}\) from the real spatial-delay CSI \(\mathbf{X}_{t}^{s}\in\mathbb{R}^{(M^{s}_{B} \times M^{s}_{U}) \times ({K}^{s} \times 2)}\):
\begin{equation}
    \mathbf{X}_{t}^{s}{'} = \text{FFN}((\text{FFN}((\mathbf{X}_{t}^{s})^{\mathsf T}))^{\mathsf T}),
\end{equation}
\begin{equation}
    \text{FFN}(\mathbf{X}) = (\text{ReLU}(\mathbf{X} \mathbf{W}_1 + \mathbf{b}_1)) \mathbf{W}_2 + \mathbf{b}_{2},
\end{equation}
where \(\mathbf{X}_{t}^{s}\) can be obtained via concatenating the real and imaginary parts of the original complex channel matrix \(\hat{\mathbf{H}}^s_t\), and \(\text{ReLU}(\cdot)\) denotes the Rectified Linear Unit (ReLU) nonlinear activation function. The matrices \(\mathbf{W}_1\in\mathbb{R}^{d_i \times d_h}\) and \(\mathbf{W}_2\in\mathbb{R}^{d_h \times d_o}\), along with the bias vectors \(\mathbf{b}_1\in\mathbb{R}^{d_h}\) and \(\mathbf{b}_2\in\mathbb{R}^{d_o}\), are learnable parameters. Here, \(d_i\), \(d_h\), and \(d_o\) denote the input, hidden, and output dimensions of the FFN, respectively.\footnote{Unless otherwise specified, these dimensions follow the relationship \(d_i = d_o = (1/2)d_h\) throughout the paper.}

\textbf{MDFM}:
An MDFM is designed to efficiently fuse the spatial, frequency, and delay domain features of the sub-6 GHz CSI. To accelerate the convergence of model training, the rearranged CSI is first normalized using the mean and standard deviation computed over each batch. Then, the normalized CSI is linearly projected into a latent space of dimension \(d_{re}\) through a dense layer \(\mathbf{W}_{re}\) to unify the size of the model and prevent exponential growth in computational complexity:
\begin{equation}
\mathbf{X}_{f}^{s}{'} = \mathbf{X}_{f}^{s} \mathbf{W}_{re},
\end{equation}
where \(\mathbf{X}_{f}^{s}\in\mathbb{R}^{(M^{s}_{B} \times M^{s}_{U})\times(K^{s}\times 2)}\) and \(\mathbf{X}_{f}^{s}{'}\in\mathbb{R}^{(M^{s}_{B} \times M^{s}_{U})\times d_{re}}\) denote the normalized spatial-frequency CSI and spatial-frequency representation, respectively, and \(\mathbf{W}_{re}\in\mathbb{R}^{(K^{s}\times 2)\times d_{re}}\) is the learnable projection matrix.
To capture the feature correlations in the spatial-frequency domain of the sub-6 GHz CSI and enhance the model's representational capacity, we employ a MHSA layer, as depicted in Fig. \ref{fig:MHSA}. However, since the MHSA mechanism will inherently ignore positional relationships, a learnable positional embedding \(\mathbf{P} \in \mathbb{R}^{(M^{s}_{B} \times M^{s}_{U}) \times  d_{re}}\) is first introduced to adaptively provide unique positional information to different antenna elements. Specifically, the input of MHSA, denoted as \(\mathbf{X}_{\text{PE}}\in \mathbb{R}^{(M^{s}_{B} \times M^{s}_{U}) \times  d_{re}}\), can be obtained as:
\begin{equation}
\mathbf{X}_{\text{PE}} = \mathbf{X}^{s}_{f}{'}+\mathbf{P},
\end{equation}
where \(\mathbf{P}\) is initialized with values drawn from a standard normal distribution \(\mathcal{N}(0, 1)\). Then, the output of each attention head is computed as:
\begin{equation}
\mathbf{O}_i = \text{Attention}(\mathbf{Q}_i, \mathbf{K}_i, \mathbf{V}_i), \quad i = 1, 2, \dots, N_h,
\end{equation}
where $\text{Attention}(\cdot)$ represents the scaled dot-product attention function with the query matrix $\mathbf{Q}_i \in \mathbb{R}^{(M^{s}_{B} \times M^{s}_{U}) \times d_k}$, the key matrix $\mathbf{K}_i \in \mathbb{R}^{(M^{s}_{B} \times M^{s}_{U}) \times d_k}$, and the value matrix $\mathbf{V}_i \in \mathbb{R}^{(M^{s}_{B} \times M^{s}_{U}) \times d_v}$. The matrices \(\mathbf{Q}_i\), \(\mathbf{K}_i\), and \(\mathbf{V}_i\) are linearly projected from the input \(\mathbf{X}_{\text{PE}}\) , with $d_k = d_v = d_{re}/N_h$, where $N_h$ denotes the number of attention heads. Then, the outputs of all attention heads are concatenated and linearly projected to obtain the output $\mathbf{Y}_a \in \mathbb{R}^{(M^{s}_{B} \times M^{s}_{U}) \times d_{{re}}}$:
\begin{equation}
\mathbf{Y}_a = [\mathbf{O}_1, \dots, \mathbf{O}_{N_h}] \mathbf{W}_o,
\end{equation}
where \(\mathbf{W}_o\in\mathbb{R}^{d_{re}\times d_{re}}\) denotes the output projection matrix.
Subsequently, the spatial, frequency, and delay domain characteristics are fused within the MoE layer (Fig. \ref{fig:MoE}). Specifically, the gating network receives the latent representation \( \mathbf{X}_{t}^{s}{'}\) from the TFEM and produces gating matrix \(\mathbf{G}\in\mathbb{R}^{(M^{s}_{B} \times M^{s}_{U})\times N_e}\) for \( N_e \) experts:
\begin{equation}
\mathbf{G} = \mathbf{X}_{t}^{s}{'} \mathbf{W}_G + \mathbf{b}_G.
\end{equation}
Then, for each row of \(\mathbf{G}\), only the \(K\) largest gating values are preserved while the remaining entries are set to zero:
\begin{equation}
    \mathbf{\tilde{G}}_{[i,j]} =
    \begin{cases}
    1, & j \in {\text{argTopK}} (\mathbf{G}_{[i,:]})\\
    0, & \text{otherwise}
    \end{cases}, i \in \{1,...,(M^{s}_{B} \times M^{s}_{U}) \},
\end{equation}
\begin{equation}
\mathbf{G'} = \text{softmax}(\mathbf{\tilde{G}}\odot\mathbf{G}),
\end{equation}
where \( \odot \) denotes the Hadamard (element-wise) product, \( \mathbf{\tilde{G}} \in \mathbb{R}^{(M^{s}_{B} \times M^{s}_{U})\times N_e}\) denotes the masked logits matrix, \(\mathbf{G'}\) represents the final gating matrix for $K$ selected experts, and $\text{softmax}(\cdot)$ denotes the SoftMax activation function. Given \(\mathbf{Y}_a\), the output of the $j$-th expert \(\mathbf{E}_j\in \mathbb{R}^{(M^{s}_{B} \times M^{s}_{U}) \times d_{re}}\) is expressed as:
\begin{equation}
\mathbf{E}_j = \text{FFN}(\mathbf{Y}_a),
\end{equation}
where the hidden dimension of each expert is \(d_e=d_{hid}/N_e\). The final output of the MoE layer \( \mathbf{Y}_e \in \mathbb{R}^{(M^{s}_{B} \times M^{s}_{U}) \times d_{re}} \) is a weighted sum of expert outputs:
\begin{equation}
\mathbf{Y}_e = \textstyle\sum_{j=1}^{N_e} \mathbf{G'}_{[:,j]} \odot \mathbf{E}_j.
\end{equation}

The sparse activation strategy of the MoE layer effectively reduces computational complexity while maintaining model performance. This design enables spatial-frequency CSI to be dynamically routed by spatial-delay information, thereby enhancing the efficiency of feature extraction and fusion.

\textbf{DFIM}:
To process the multi-domain features extracted by the MDFM, the DFIM adopts a multi-layer deep neural network. Specifically, each layer integrates a MHSA mechanism, MoE layer, and a residual connection structure, which is similar to the MDFM. However, unlike the MDFM, which focuses on multi-domain feature extraction and fusion, the DFIM's MoE layer utilizes identical inputs for both the gating and expert networks. This design enhances channel knowledge extraction while preserving computational efficiency. The final output of the DFIM is the low-rank latent embedding of the estimated mmWave channel.

At the output stage of MDFCE, a linear mapping layer followed by denormalization is employed to project the latent embedding back to the spatial-frequency domain. The offline training objective is to minimize the normalized NMSE between the estimated and ground truth mmWave CSI:
\begin{equation}
\mathcal{L}_{\text{NMSE}} = 1/N \textstyle\sum_{i=1}^{N} \| \mathbf{H}^{m}_{i} - \hat{\mathbf{H}}^{m}_{i} \|_2^2 / \| \mathbf{H}^{m}_{i} \|_2^2,
\end{equation}
where ${{\mathbf{{H}}}^{m}_{i}}$ and ${\hat{\mathbf{{H}}}^{m}_{i}}$ are the ground truth and the estimated mmWave CSI, respectively, and \(N\) is the number of training samples. 

To prevent routing collapse and promote balanced utilization of all experts, we adopt a load-balancing auxiliary loss $\mathcal{L}_{{aux}}$ \cite{lepikhin2021gshard}. Specifically, the mean gating value and routing fraction of the $j$-th expert are defined as $m_j = 1/M^s \textstyle\sum_{i=1}^{M^s} \mathbf{G'}_{[i,j]} $ and $p_j = 1/M^s \textstyle\sum_{i=1}^{M^s} \mathbf{\tilde{G}}_{[i,j]}$, respectively, where $M^s=M^{s}_{B} \times M^{s}_{U}$. Then $\mathcal{L}_{{aux}}$ is formulated as:
\begin{equation}
\mathcal{L}_{{aux}} = N_e \textstyle\sum_{j=1}^{N_e} m_j p_j,
\end{equation}
and the total loss function is defined as:
\begin{equation}
\mathcal{L}_{{total}} = \kappa * \mathcal{L}_{\text{NMSE}} + (1-\kappa) * \mathcal{L}_{{aux}},
\end{equation}
where $\kappa$ is a loss balancing hyper-parameter. 

In order to improve training stability and convergence, residual connections and layer normalization are employed in the module, while related expressions are omitted for brevity.
\section{Experiments}
In this section, we adopt the DeepMIMO dataset \cite{alkhateeb2019deepmimo} to evaluate the effectiveness of the proposed MDFCE. First, we describe the experimental settings in detail. Then, we compare the MDFCE with the conventional pilot-based LS channel estimation method \cite{504981} to verify the superiority of cross-band channel extrapolation over traditional pilot-based channel estimation. Finally, ablation experiments demonstrate the performance gains brought by spatial-delay-frequency domain feature fusion. Comparisons with the TBN \cite{10024904} further highlight the advantages of the MoE architecture in both extrapolation performance and computational efficiency. In addition, comparisons with the latest MCBAM \cite{li2024physicsenabled}, which leverages sub-6 GHz CSI together with partial mmWave pilots for mmWave channel extrapolation, further verify the superiority of the proposed MDFCE.
\subsection{Experimental Settings}
The outdoor scenario `O1' in the DeepMIMO dataset \cite{alkhateeb2019deepmimo} is adopted.
The uplink works at 3.5 GHz and the downlink works at 28 GHz. We select `BS2' as the base station, and locations from rows 250 to 749, with each row containing 181 locations, as UE locations, yielding a total of 90,500 samples. The dataset is divided into training and validation sets with a ratio of 7:3. Detailed system and hyper-parameter settings are given in Table~\ref{tab:exp-settings}. Normalized mean square error (NMSE) in decibels (dB) is adopted to evaluate the channel extrapolation performance.

\begin{table}[t]
    \centering
    \caption{System and hyper-parameter settings.}
    \setlength{\tabcolsep}{3.5pt}
    \begin{tabular}{c|c|c|c}
    \hline
    \multicolumn{4}{c}{\textbf{System settings}} \\
    \hline
    \multicolumn{2}{c|}{ITEM} & \multicolumn{2}{c}{SETTINGS} \\
    \hline
    \multicolumn{2}{c|}{Freq. bands (UL/DL)} & \multicolumn{2}{c}{3.5 GHz / 28 GHz} \\
    \hline
    \multicolumn{2}{c|}{UE antennas (UL/DL)} & \multicolumn{2}{c}{2 / 2} \\
    \hline
    \multicolumn{2}{c|}{BS antennas (UL/DL)} & \multicolumn{2}{c}{4,16 / 8,32} \\
    \hline
    \multicolumn{2}{c|}{Spacing (UL/DL)} & \multicolumn{2}{c}{0.5$\lambda$ / 0.5$\lambda$} \\
    \hline
    \multicolumn{2}{c|}{Bandwidth (UL/DL)} & \multicolumn{2}{c}{40 MHz / 123 MHz} \\
    \hline
    \multicolumn{2}{c|}{Subcarriers (UL/DL)} & \multicolumn{2}{c}{128 / 256} \\
    \hline
    \multicolumn{2}{c|}{Paths (UL/DL)} & \multicolumn{2}{c}{15 / 5} \\
    \hline
    \multicolumn{4}{c}{\textbf{Hyper-parameter settings}} \\
    \hline
    ITEM & SETTINGS & ITEM & SETTINGS \\
    \hline
    Learning rate & $1e-4$ & Epochs & $1000$ \\
    \hline
    Batch size & 128 & Optimizer & AdamW \\
    \hline
    DFIM blocks $N_{s}$ & 7 & Heads $N_{h}$ & 4 \\
    \hline
    Rep. dim. $d_{re}$ & 128 & Hidden dim. $d_{hid}$ & 256 \\
    \hline
    Experts $N_e$ & 8 & Expert dim. $d_{e}$ & $d_{hid}/N_e=32$ \\
    \hline
    Top-K experts $K$ & 2 & Balance factor $\kappa$ & 0.99 \\
    \hline 
    \end{tabular}
    \label{tab:exp-settings}
    \vspace{-4mm}
\end{table} 
\begin{figure}[!t]
      \centering
      \includegraphics[width=0.7\linewidth]{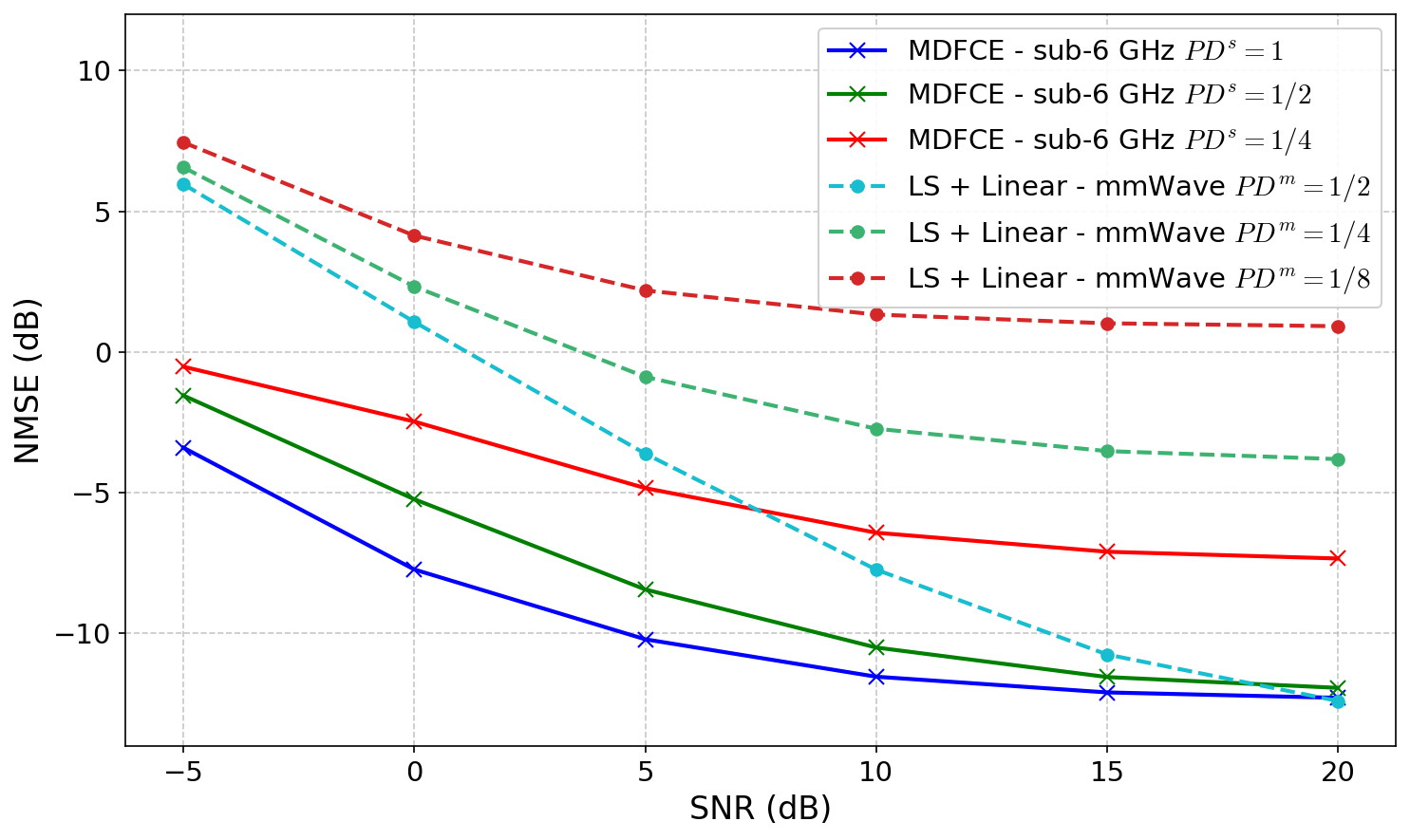}
      \caption{Comparison of pilot-based direct mmWave channel estimation and cross-band channel extrapolation.}
      \label{fig:pilot-compare}
      \vspace{-8mm}
  \end{figure}


\subsection{Effectiveness of the Cross-Band Channel Extrapolation}
We first compare the performance of dual-band channel extrapolation based on the proposed MDFCE with uplink pilot-based direct mmWave channel estimation using the LS method with linear interpolation. Here we assume that the system operates in time division duplexing (TDD) mode, so the downlink mmWave CSI can be obtained via uplink channel estimation to facilitate pilot-overhead comparison. It is worth emphasizing that the proposed method is applicable to both TDD and frequency division duplexing (FDD) systems, and it is expected to offer greater advantages in FDD systems, where downlink channel estimation and CSI feedback incur significantly higher overhead than uplink channel estimation. Pilots are uniformly placed across all subcarriers with frequency-domain pilot density (PD) defined as $PD^s=K^s_{\mathrm{pilot}}/K^s$ for the sub-6 GHz band and $PD^m=K^m_{\mathrm{pilot}}/K^m$ for the mmWave band, where $K^s_{\mathrm{pilot}}$ and $K^m_{\mathrm{pilot}}$ denote the numbers of sub-6 GHz and mmWave subcarriers carrying pilots, respectively. For the dual-band channel extrapolation scheme, the sub-6 GHz array at the UE and BS is equipped with 2 and 16 antennas, respectively. The sub-6 GHz UE transmits pilots to estimate the sub-6 GHz CSI, which is then used to extrapolate the mmWave CSI via the proposed MDFCE. For the direct mmWave channel estimation scheme, the UE and BS are equipped with 2 and 32 mmWave antennas, respectively, and the mmWave UE transmits pilots for channel estimation at the BS using the LS method. The pilot overhead for dual-band channel extrapolation is $2 \times 128 \times PD^s$, while the pilot overhead for direct mmWave channel estimation is $2 \times 256 \times PD^m$. As shown in Fig. \ref{fig:pilot-compare}, even under low SNR, the NMSE of the proposed MDFCE is significantly lower than that of the pilot-based method under comparable pilot overhead (e.g., ``MDFCE - sub-6 GHz $PD^s$=1'' versus ``LS + Linear - mmWave $PD^m$=1/2''). In addition, ``MDFCE - sub-6 GHz $PD^s$=1/4'' achieves a $50\%$ reduction in pilot overhead and an average 4.44 dB performance improvement compared with ``LS + Linear - mmWave $PD^m$=1/4''. These results demonstrate the superiority of cross-band channel estimation in terms of noise robustness, pilot overhead, and mmWave channel estimation accuracy.
  \begin{figure}[!t]
      \centering
      \includegraphics[width=0.6\linewidth]{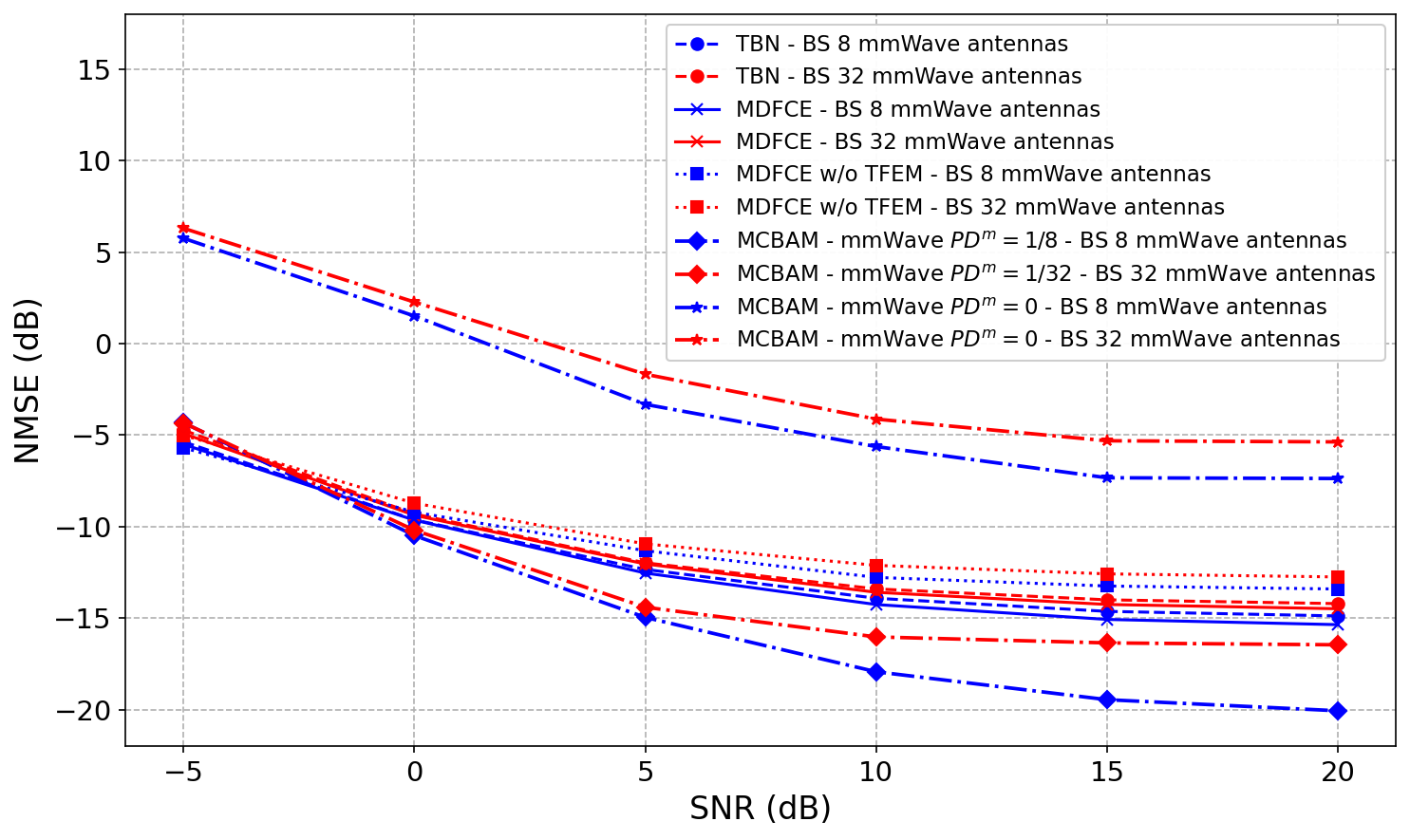}
      \caption{Comparison of NMSE for the MDFCE, the MDFCE without TFEM module, the TBN, and the MCBAM.}
      \label{fig:Others_vs_Ours}
      \vspace{-6mm}
  \end{figure}

  \begin{figure}[!t]
      \centering
      \includegraphics[width=0.6\linewidth]{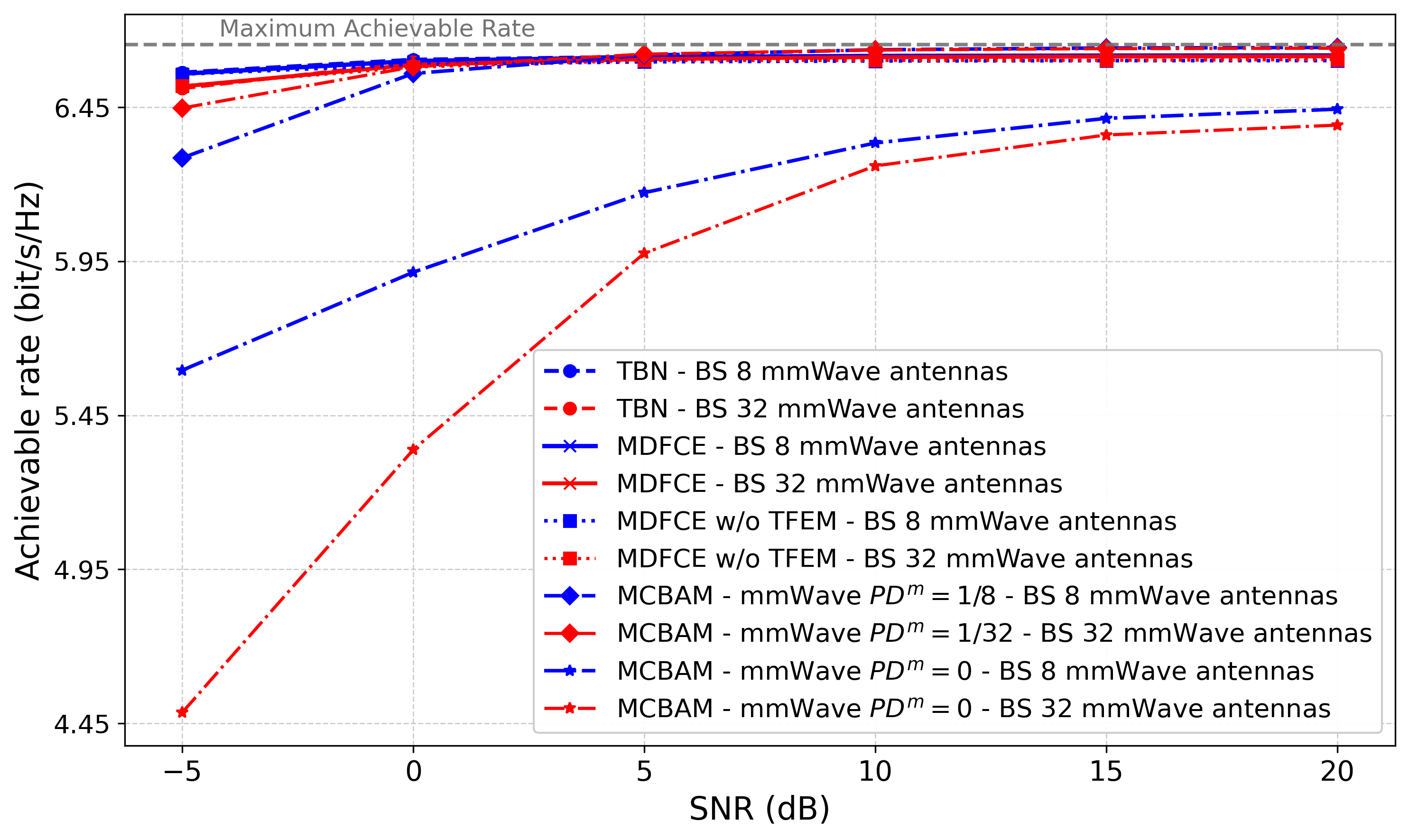}
      \caption{Comparison of achievable rate for the MDFCE, the MDFCE without TFEM module, the TBN, and the MCBAM.}
      \label{fig:SE_Comparison}
      \vspace{-8mm}
  \end{figure}
\subsection{Effectiveness of the MDFCE}
Fig. \ref{fig:Others_vs_Ours} presents a comparative analysis of NMSE versus SNR for the MDFCE, the MDFCE without TFEM module, the TBN, and the MCBAM. The UE in both bands employs 2 antennas, while the BS employs 4 sub-6 GHz antennas. The TBN and MCBAM are implemented following \cite{10024904} and \cite{li2024physicsenabled}, respectively, in which MCBAM requires partial mmWave CSI as additional input. For a fair comparison, the model sizes of the TBN and MCBAM baselines are configured to be comparable to that of the proposed MDFCE. Compared with the MDFCE without TFEM module, the MDFCE showcases an average performance gain of 1.1 dB when the BS has 8 or 32 mmWave antennas. This result demonstrates that leveraging the spatial-delay domain features extracted by the TFEM module from the sub-6 GHz channel to guide the fusion of the spatial-frequency features extracted by the MDFM module in the MoE layer enables better cross-band channel knowledge learning and more accurate mapping learning. In addition, under the 32-mmWave-antenna configuration, the MCBAM requires mmWave CSI with $PD^m$=1/32 as additional input to achieve performance comparable to the proposed MDFCE. When mmWave CSI is extrapolated solely from sub-6 GHz CSI (i.e., $PD^m$=0), the MDFCE outperforms the MCBAM by 9.64 dB, demonstrating that the proposed method achieves superior performance with lower pilot overhead. Compared to the TBN, which employs a more complex network architecture with the same input, the proposed MDFCE achieves comparable or even superior performance. Furthermore, Fig. \ref{fig:SE_Comparison} compares the achievable rate of different methods, where the extrapolated mmWave CSI is used for maximum ratio transmission (MRT) precoding and the rate is evaluated over the ground-truth mmWave channel. The MDFCE obtains an achievable rate of 6.60 bit/s/Hz, which is identical to that of the TBN and 0.63 bit/s/Hz higher than that of the MCBAM. Moreover, the MCBAM requires additional mmWave channel information equivalent to 1/4 of that in the sub-6 GHz band to achieve an achievable rate of 6.58 bit/s/Hz, whereas the MDFCE achieves comparable or even better performance with lower pilot overhead. This demonstrates that the proposed MDFCE can more efficiently reconstruct the precise amplitude and phase information necessary for digital precoding compared with the TBN and MCBAM.

Owing to the sophisticated network design, additional spatial-delay domain feature fusion, and computationally efficient MoE architecture, the MDFCE greatly reduces network complexity and improves inference speed while preserving high performance, making it more promising for practical deployment in communication systems. Specifically, the MDFCE achieves approximately \textbf{1.33$\times$} inference speedup and a \textbf{2.42$\times$} reduction in floating-point operations (FLOPs) per sample, compared to the TBN, on an NVIDIA RTX 3090 GPU. With a batch size of 128, the TBN requires \textbf{0.262 ms} and \textbf{2.72 GFLOPs} per sample for inference. In contrast, our method reduces both the inference time (to \textbf{0.197 ms}) and computational cost (to \textbf{1.12 GFLOPs}), while maintaining or even outperforming the TBN in NMSE performance. Compared to the proposed MDFCE, the lightweight MCBAM delivers an \textbf{8.6$\times$} inference speedup and a \textbf{18.8$\times$} reduction in FLOPs, with only \textbf{0.06 GFLOPs} per sample, but exhibits a pronounced performance degradation when mmWave band information is unavailable.
\vspace{-3mm}
\section{Conclusion}
In this work, we proposed the MDFCE, a novel deep learning-based architecture for mmWave CSI acquisition via cross-band channel extrapolation. By leveraging a gating mechanism inspired by the MoE framework, the MDFCE effectively fused spatial, frequency, and delay domain features of wireless channels, addressing the highly nonlinear and intractable mapping between sub-6 GHz and mmWave channels. Extensive evaluations on the DeepMIMO dataset under varying antenna array sizes and SNR levels demonstrated that MDFCE outperformed conventional methods in terms of channel estimation accuracy, pilot overhead, and computational efficiency.
\bibliographystyle{IEEEtran} 
\bibliography{Zhou_WCL2026-1045_refs}
\end{document}